\documentclass[sigconf]{acmart}
\def\BibTeX{{\rm B\kern-.05em{\sc i\kern-.025em b}\kern-.08emT\kern-.1667em\lower.7ex\hbox{E}\kern-.125emX}}

\acmYear{2019}
\copyrightyear{2019}
\setcopyright{acmcopyright}
\acmConference[SocialSense '19]{SocialSense '19: Fourth International Workshop on Social Sensing}{April 15, 2019}{Montreal, QC, Canada}
\acmBooktitle{SocialSense '19: Fourth International Workshop on Social Sensing, April 15, 2019, Montreal, QC, Canada}
\acmPrice{15.00}
\acmDOI{10.1145/3313294.3313385}
\acmISBN{978-1-4503-6706-6/19/04}

\begin{document}

%
% The "title" command has an optional parameter, allowing the author to define a "short title" to be used in page headers.
%\title[CarbonKit]{CarbonKit:\\A Technological Platform for Personal Carbon Tracking}
\title[CarbonKit]{CarbonKit: Designing A Personal Carbon Tracking Platform}

%
% The "author" command and its associated commands are used to define the authors and their affiliations.
% Of note is the shared affiliation of the first two authors, and the "authornote" and "authornotemark" commands
% used to denote shared contribution to the research.

\author{Laura Guzman}
\affiliation{%
 \institution{Simon Fraser University}
 \city{Burnaby}
 \state{BC}
 \country{Canada}}
 \email{lguzmanf@sfu.ca}

\author{Stephen Makonin}
\affiliation{%
 \institution{Simon Fraser University}
 \city{Burnaby}
 \state{BC}
 \country{Canada}}
 \email{smakonin@sfu.ca}

\author{Alex Clapp}
\affiliation{%
 \institution{Simon Fraser University}
 \city{Burnaby}
 \state{BC}
 \country{Canada}}
 \email{aclapp@sfu.ca}

%
% By default, the full list of authors will be used in the page headers. Often, this list is too long, and will overlap
% other information printed in the page headers. This command allows the author to define a more concise list
% of authors' names for this purpose.
\renewcommand{\shortauthors}{Guzman, Makonin, and Clapp}

%
% The abstract is a short summary of the work to be presented in the article.
\begin{abstract}
Ubiquitous technology platforms have been created to track and improve health and fitness; similar technologies can help individuals monitor and reduce their carbon footprints. This paper proposes CarbonKit -- a platform combining technology, markets, and incentives to empower and reward people for reducing their carbon footprint. We argue that a goal-and-reward behavioural feedback loop can be combined with the Big Data available from tracked activities, apps, and social media to make CarbonKit an integral part of individuals\rq{} daily lives. CarbonKit comprises five modules that link personal carbon tracking, health and fitness, social media, and economic incentives. Protocols for safeguarding security, privacy and individuals\rq{} control over their own data are essential to the design of the CarbonKit. We use the example of the British Columbia to illustrate the regulatory framework and participating stakeholders that would be required to implement the CarbonKit in specific jurisdictions.
\end{abstract}

%
% The code below is generated by the tool at http://dl.acm.org/ccs.cfm.
% Please copy and paste the code instead of the example below.
%
\begin{CCSXML}
<ccs2012>
<concept>
<concept_id>10003120.10003138.10003139.10010904</concept_id>
<concept_desc>Human-centered computing~Ubiquitous computing</concept_desc>
<concept_significance>500</concept_significance>
</concept>
<concept>
<concept_id>10003456.10003457.10003458.10010921</concept_id>
<concept_desc>Social and professional topics~Sustainability</concept_desc>
<concept_significance>500</concept_significance>
</concept>
</ccs2012>
\end{CCSXML}

\ccsdesc[500]{Human-centered computing~Ubiquitous computing}
\ccsdesc[500]{Social and professional topics~Sustainability}

%
% Keywords. The author(s) should pick words that accurately describe the work being
% presented. Separate the keywords with commas.
\keywords{behavioral change, big data, carbon footprint, personal carbon tracking, computational sustainability, ubiquitous platform}

%
% A "teaser" image appears between the author and affiliation information and the body
% of the document, and typically spans the page.
\begin{teaserfigure}
  \includegraphics[width=\textwidth]{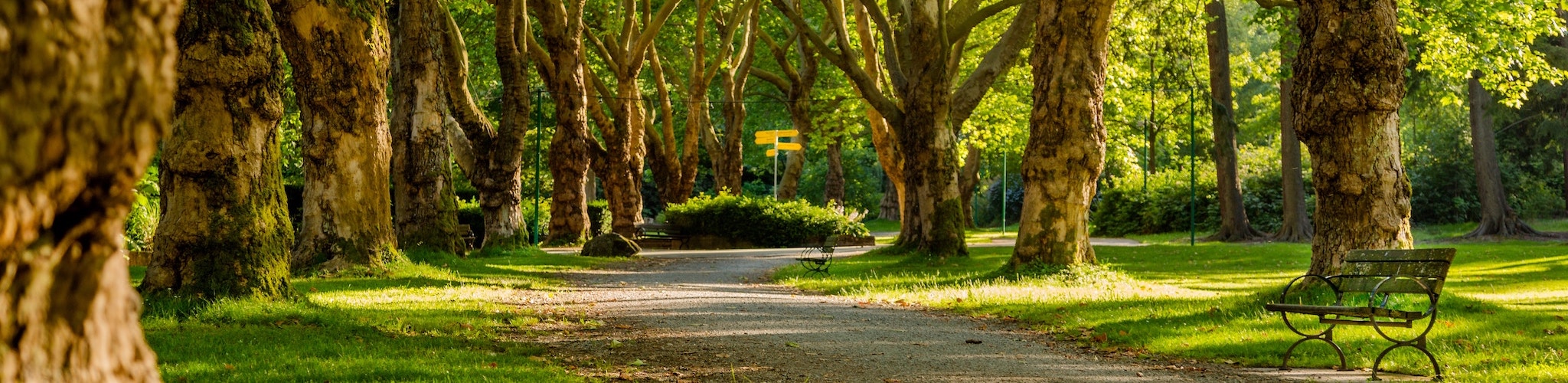}
  \caption{Stanley Park, Vancouver, Canada. Photograph by Mike Benna on Unsplash}
  \label{fig:teaser}
\end{teaserfigure}

%
% This command processes the author and affiliation and title information and builds
% the first part of the formatted document.
\maketitle

\section{Introduction}

The 1992 United Nations Framework Convention on Climate Change committed signatory states to reduce greenhouse gas (GHG) emissions, based on the premises that global warming was a reality, and that anthropogenic emissions were the primary cause~\cite{unfccc1992united}. Despite the agreement, from 1992 to 2015, global GHG emissions increased over 35 percent. Estimated worldwide Carbon Dioxide (CO2) emissions for 2015 totalled nearly 35.7 billion metric tons. During this period, numerous policies have been proposed and some of them implemented, including carbon taxes, emissions trading systems, and low carbon fuel standards, as well as advancing research and development of clean technologies.

The 21st Conference of the Parties (COP21) in Paris reached a non-binding agreement committing 195 Nations to reduce their carbon output and to limit global warming to a maximum of 1.5\textcelsius~\cite{rogelj2016paris}. This paper begins from the premise that industry-focused policies alone will not suffice to limit warming to 1.5\textcelsius: emissions reductions will be needed from all sectors of the economy, including individuals. Transition to a sustainable society will require shifts in consumer lifestyles as well as the development of new technologies~\cite{grubb2015climate}. This paper explores how technology and incentives might be combined in ways that reward and empower individuals\rq{} efforts to reduce GHG emissions.

This mass coordination of public and private companies, governments, and people would require a computing platform that is pervasive and ubiquitous so that data used to track carbon consumption is gathered with minimal manual effort. Incentives can be offered in real-time based on geo-location or current activity. A mix of sensors and near field communication (NFC) devices would form the basis of interaction with such a pervasive computing platform. Sensors and NFC devices would track, for example, the amount of distance travelled by transit and the amount of gas used by car, compare the two, and offer an incentive -- show savings -- when transit is used.

We propose a CarbonKit -- a combination of computing applications, information sources, and incentives that would enable individuals to track and reduce their personal carbon emissions. It combines economic, psychological and social incentives for individual engagement and behavioral change with a stakeholder model. This paper first describes the technology platform required for the CarbonKit, and then explains how the CarbonKit would address privacy and security. Implementing a CarbonKit would require a regionally specific network of stakeholders. Using the Canadian province of British Columbia (BC) as an example, this paper identifies a network of public and private institutions whose core competencies and functions would be foundational for the CarbonKit\rq{}s design, implementation and operation.

\section{Technology \& Behavioral Change}

Behavioral change is a significant and consistent change in the way and the frequency an individual uses technologies, institutions and infrastructure. Although behavioral change initially requires conscious effort, once institutionalized and habituated, the new behavior can become effortless, unconscious and durable over time. Initiating behavioral change is an uphill battle because individuals find changes to routine difficult and cognitively draining. Habits are difficult to break because the formation of a new habit uses cognitive energy, while maintaining an existing habit is automatic~\cite{ariely2009doing}. Nevertheless, some behaviors can be modified by changes in the built environment in which individuals live -- such as higher-density communities or closer proximity to transit -- or by providing smart technological solutions that reduce the need for cognitive effort and help improve social and environmental conditions for individuals.

The less cognitive energy and time that an individual requires for a task, the greater the potential to modify the way that activity is performed. How much technology can influence behavior over the long term is an open research question in the field of human-computer interaction. This paper argues that if technology is ubiquitous and feedback is readily available, then its influence on behavior is greater, particularly if the feedback is compelling and creates a sense of or a call to action~\cite{makonin2013smarter}.

Identifying barriers to behavioral change is also a crucial step for technology development to mitigate climate change. Gifford identified 30 different psychological barriers to changing environmental behavior, of which the three most significant were conflicting goals, perceived lack of efficacy and social comparison~\cite{gifford2011dragons}. As social beings, humans regularly compare themselves to others to better understand who they are, how they fit in communities, and what behavior is appropriate to a particular situation~\cite{guimond2006social}. Seeing peers discount the threat of climate change can have a powerful effect on people, discouraging them from acting as well; conversely, social comparison with peers who have reduced their carbon footprint can be a powerful motivator.

Related barriers to be considered in technology design are the human focus on tangible risks and the potential for information overload. Individuals are hardwired to respond to tangible and immediate risks, and to discount the impact of future or distant risks~\cite{team2011behaviour}. Behavioral change requires the major, conscious effort usually given to urgent or engaging matters, such as when important aspects of their daily lives are affected. If a software application for carbon tracking and reduction is also linked to short-term human concerns and goals, it is more likely to promote long-term change. Health, sports, economy, and social recognition and cohesion are all highly relevant in people\rq{}s daily lives, and are desirable near-term co-benefits~\cite{guzman2017applying}.

Strategies for promoting behavioral change include default options, social proofing, gamification, and incentives. Individuals tend to choose the default, automatic, or most available behavior: if a green choice is the default option, and the more carbon-intensive choice requires an opt-out action, people will be more likely to choose the green option~\cite{gifford2014environmental}. Social proofing also takes advantage of social norms, as when an individual looks to others for appropriate behavior to imitate. The more people who exhibit an observed behavior, the more likely that someone will model that behavior. If individuals believe that most of their neighbours, coworkers or friends are adopting a behavior, they are more likely to do the same~\cite{naumof2013values}. Showing people when and how their neighbours (or circle of influence) are ``doing the right thing'' can promote a desired behavior. Communication of desired behaviors and the reasons for them work best through channels that have already proven to influence people, such as social networks and new media.

Gamification techniques seek to leverage people\rq{}s natural desires for competition, achievement, status, self-expression, altruism, and closure~\cite{vine2016competition}. A core gamification strategy rewards players who accomplish desired tasks. Types of rewards include points, achievement badges or levels, the filling of a progress bar, or providing the user with virtual (alternative) currency: successful games are based around discovery and accomplishment. Fitbit, Misfit and Moov are examples of applications designed to improve health and fitness that use social networks to facilitate the connection and competition among participants.

\section{The CarbonKit}

\begin{figure}[h]
  \centering
  \includegraphics[width=0.9\linewidth]{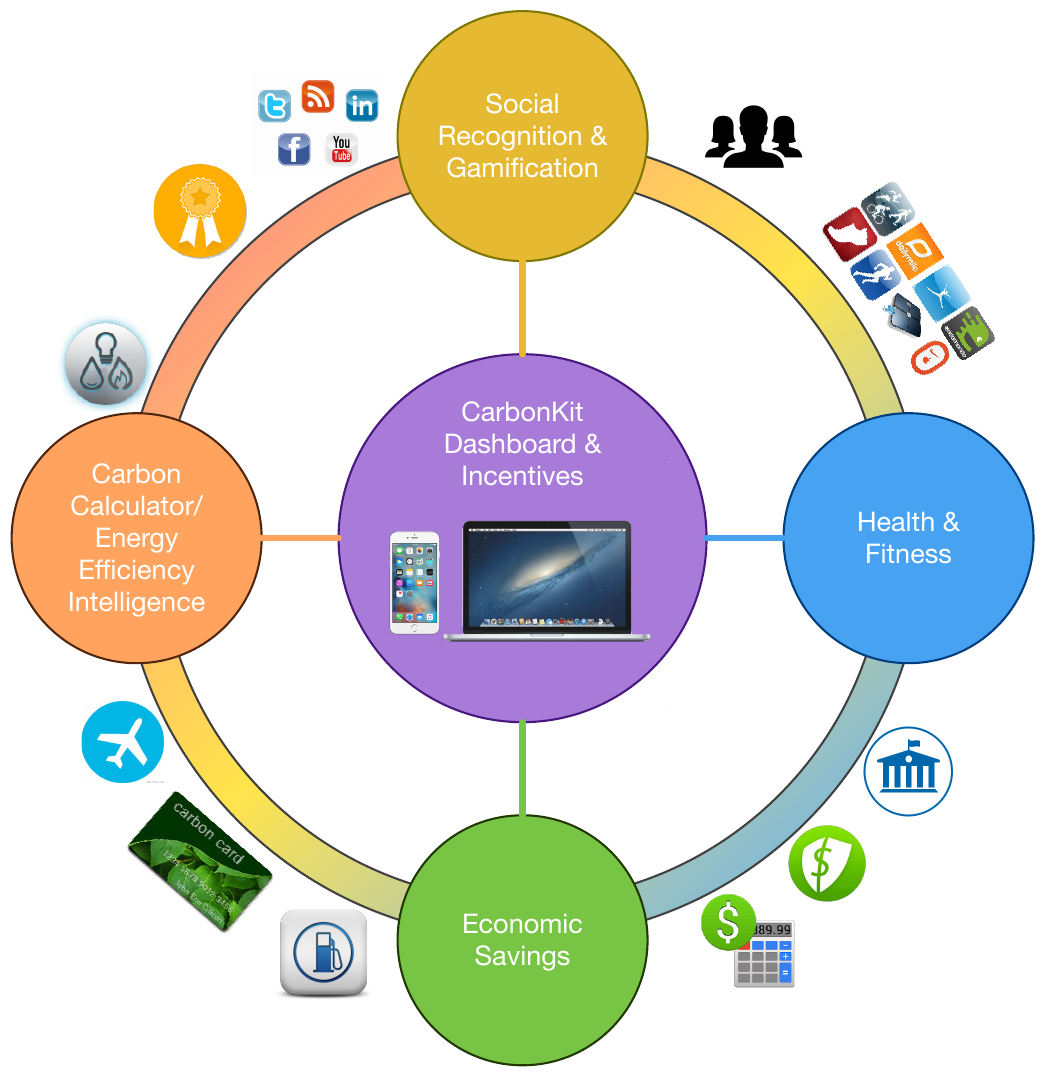}
  \caption{Overall view of the CarbonKit.}
  \label{fig:whole}
\end{figure}

The CarbonKit is envisioned as a ubiquitous platform that integrates web-based and smartphone supporting applications, including carbon footprint monitoring, budgeting, and reduction, as well as health improvement and money-saving. Figure~\ref{fig:whole} depicts the links between personal carbon tracking and more immediate concerns to promote individual engagement and collective action towards climate change mitigation and the achievement of short-term personal and social goals, including health, recreation, household economy, and social recognition and cohesion.

The CarbonKit is modeled on Apple\rq{}s HomeKit, HealthKit, and ResearchKit. HomeKit provides a platform for users to control and communicate with devices and accessions that are part the user\rq{}s home (e.g., lighting control). HealthKit allows for the sharing of health and fitness services and data (e.g., step counter). ResearchKit provides an open-source platform for medical research to use data collected from HealthKit. These applications do not have developer kits, which are synonymous with having to buy a fixed piece of hardware or a set of programming libraries — in essence developing a stand-alone products or services whose integration would be left to a third party or the end users.

A Kit is more than that: it is a broadly available and open platform for the integration of complex relationships between various stakeholders. It is a boundary object with a common structure to which different participants may attach different meanings, one sufficiently pliable to allow groups with different interests to work cooperatively~\cite{affolderbach2012environmental}. The CarbonKit would provide a ubiquitous platform that supports application and device developers, service providers, and government regulators in the common goal of enabling individuals to monitor and reduce their carbon footprints.

\subsection{Objectives \& Operations}

\begin{figure}[h]
  \centering
  \includegraphics[width=0.9\linewidth]{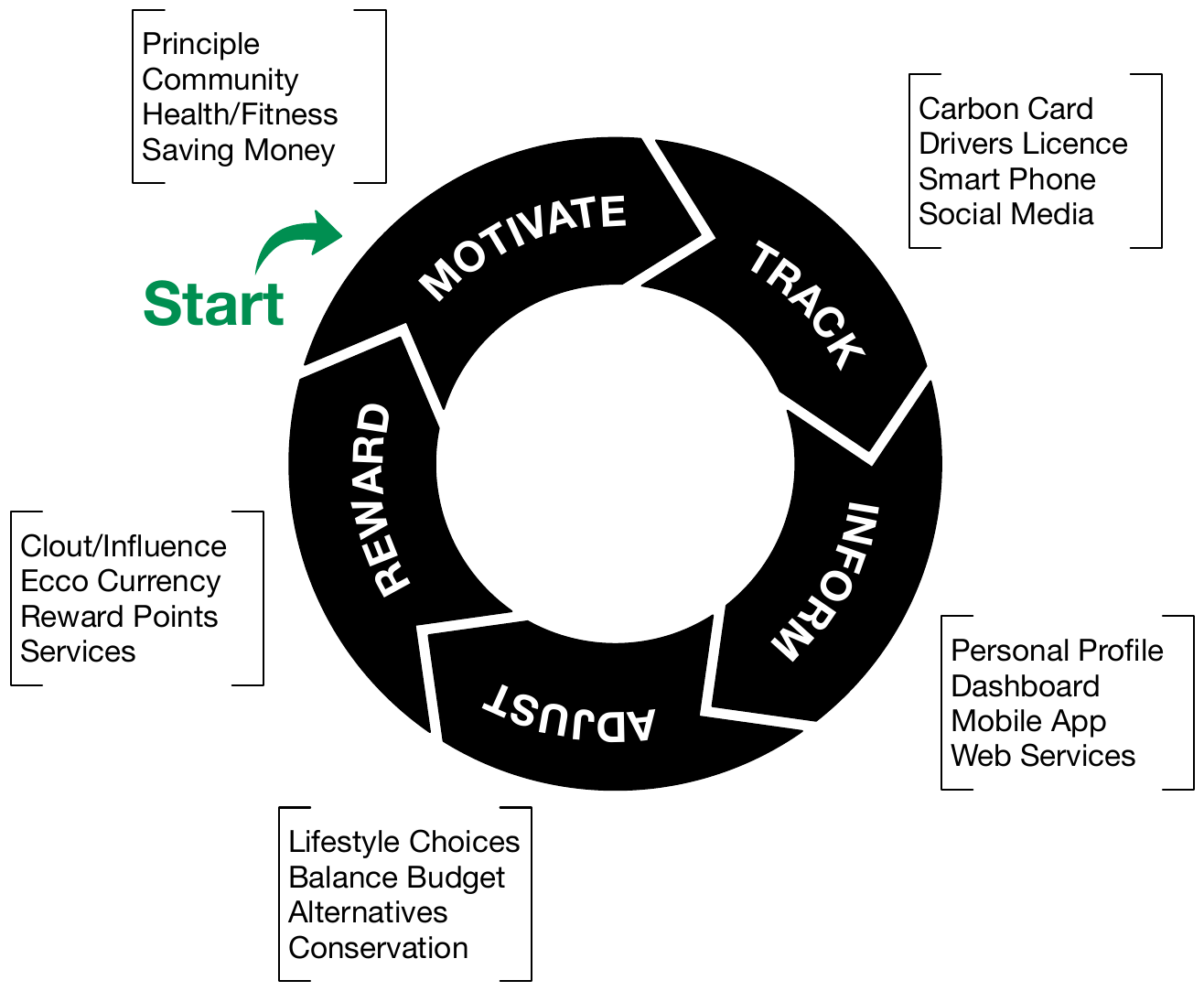}
  \caption{Behavioural feedback loop.}
  \label{fig:loop}
\end{figure}

The CarbonKit envisions a continual process of individual learning and self-improvement, at first to track and eventually to reduce personal carbon emissions. Figure~\ref{fig:loop} represents that CarbonKit platform as a behavioral change feedback loop~\cite{diclemente2001role}. Successive actions over time would aim to reduce personal carbon emissions, relative to either past emissions or to a target. Individuals\rq{} goals are likely to vary, and can be reset as their needs and preferences change over time.

Figure~\ref{fig:loop} shows the sequence of stages (motivate, track, inform, adjust, reward), with examples of each. As with most feedback loops used for behavior change, a person starts with a motivation and a clearly measurable goal. Such motivations could be internal, such as altruism, saving money, or health and fitness, or external, such as community, friends, and family, or a combination. The CarbonKit platform can track progress toward user-set goals, and social media could be used to compare an individual\rq{}s progress to friends and family~\cite{richter2015studying}.

The feedback loop closes when rewards motivate individuals to establish new goals or expand current ones. Small initial changes increase individuals\rq{} comfort level, enabling bigger changes over time.

\section{The Technology Platform}

Technology can make the complex simple by integrating solutions to apparently unrelated problems in a single platform running many applications. Through the CarbonKit dashboard, a user can access services; other linked applications vary from notification pop-ups, context menu integration, or access to any application on a mobile device or computer. Surveys suggest that the best way to deliver a new program or application is through a smartphone. About two-thirds of BC residents owned a smartphone in 2013, and more than one in four people aged 18-34 said that they cannot live without a smartphone~\cite{shaw2013phone}. It follows that the best way to implement and operate a personal carbon tracking system, or any climate change program designed for individuals, is to create a smartphone app. Many smartphone applications already help people modify their behaviors for their own, and others\rq{}, benefit. Smartphone applications offer:
\textit{portability}, accessible from any computer with an Internet connection;
\textit{mobility}, dashboard apps can be supported on diverse mobile devices (e.g., Blackberry, iPhone, or Android smartphones); and
\textit{collaboration}, individuals can access their data privately, and choose what to share with other people.

\subsection{Platform Modules}

The CarbonKit would consist of five main modules:
(1) Personal Carbon Tracking,
(2) Health and Fitness,
(3) Money Saving Tools,
(4) Social Media, and
(5) Incentives
The first three modules enable users to set goals for and track progress toward: 1) reducing carbon footprint and energy consumption; 2) improving health and fitness; and 3) saving money and adjusting to a budget. The fourth module leverages gamification and social influence to promote actions towards goals and to compare progress with peers. The fifth module manages carbon allowances and incentives earned through CarbonKit applications.

Each module would share a consistent application programming interface (API) to integrate multiple services and applications. A consistent API will help future-proof the platform to allow for new services and applications to be integrated as they are conceived and developed, thus extending the functionality of the platform.

\begin{figure}[h]
  \centering
  \includegraphics[width=0.8\linewidth]{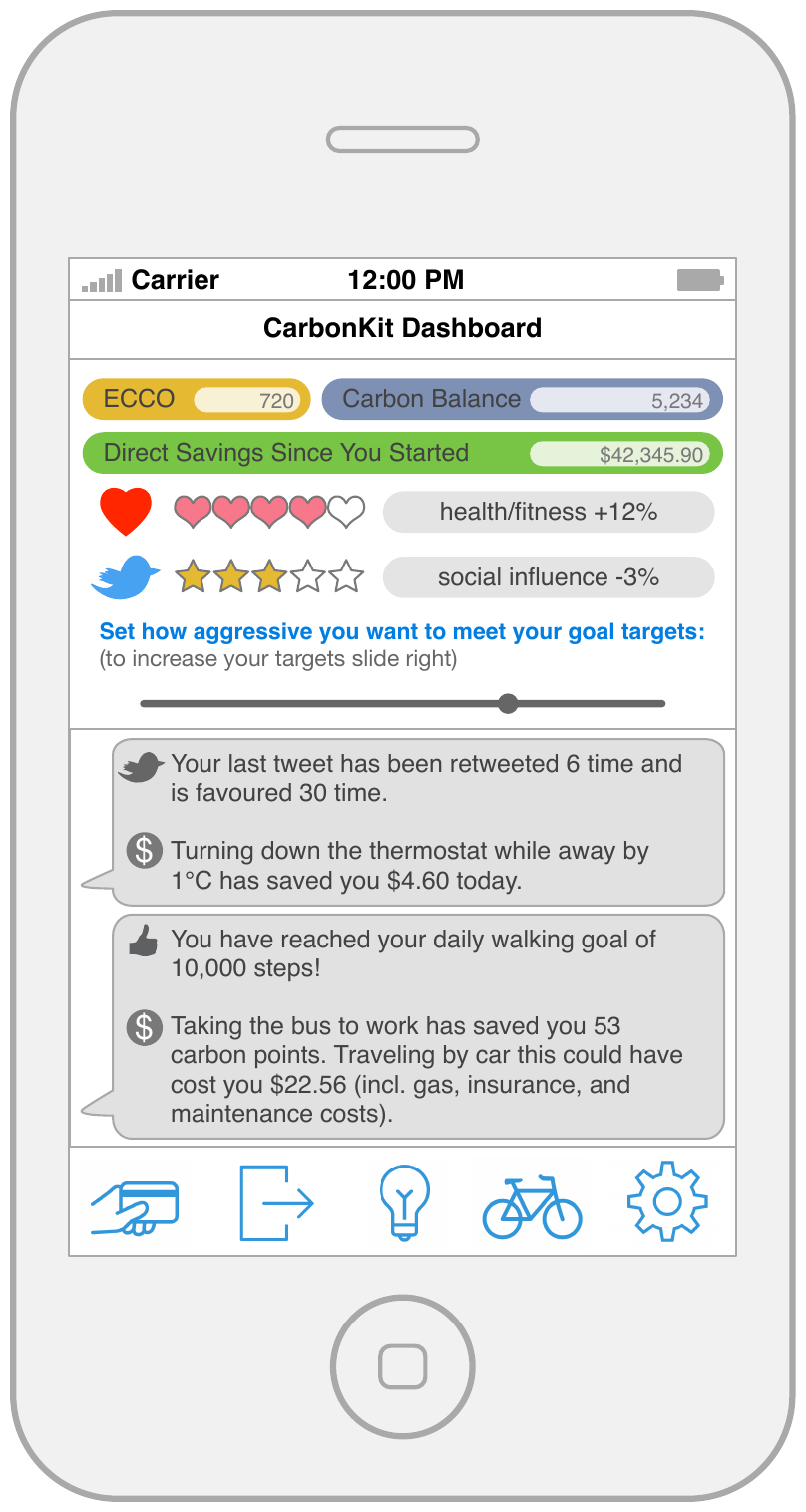}
  \caption{The CarbonKit dashboard design.}
  \label{fig:dashboard}
\end{figure}

The CarbonKit platform would allow individuals to track their progress through a personal dashboard (Figure~\ref{fig:dashboard}) on a smartphone or tablet. The CarbonKit dashboard provides a consolidated view of multiple data sources to support decision-making, tracking indicators for carbon footprint, health and fitness, economic savings, and social recognition and influence, as well as any incentives earned. It could also correlate different data sources to model how one action would impact on other indicators. As they reach goals, individuals can receive badges as further motivation~\cite{antin2011badges}. What information is displayed most prominently depends on the individual\rq{}s personal profile. The dashboard should also allow users to select their favourite interface or app in every module.

Enabling individuals to track their progress is essential to behavioral change as it allows individuals to make adjustments if they fall short of their goals. Overly ambitious goals can lead to failure and discouragement. Establishing achievable initial goals at first, and raising them later increasing them, is central to the CarbonKit platform.

After some adjustments, individuals will achieve their goals and earn rewards. Altruistic rewards might include carbon credits that can be applied toward the individual\rq{}s goals (e.g., air travel) or donated to a charity. Alternatively, individuals might collect reward points or eco-currency (an alternative currency for carbon trading) that would go toward the purchase of low-emission products or services. The CarbonKit platform could also connect consumers with businesses offering such products or services.

\subsection{Protecting Security and Privacy}

Security and privacy are vital concerns in the design of the CarbonKit platform. Although security and privacy are separate topics, they are inextricably linked. The topic of security focuses on communication and authentication between the components of the CarbonKit platform, while privacy concerns the individual users of the CarbonKit and how and by whom their information is stored and accessed. Both security and privacy affect system and data integrity, and the prevention of system intrusion, data exploitation, and unwanted data analysis.

Several design considerations can prevent system intrusion and data exploitation. Communication between apps, service providers, and the CarbonKit platform must be encrypted, and the development of the CarbonKit platform must follow secure coding practices. The encryption of stored data must also be implemented in a way that is individual-based so that individuals\rq{} data is further protected from persons with inside access to the CarbonKit platform.

Public acceptability of policies requiring the use of technology requires safeguards against privacy transgression. To ensure success and public acceptability of the CarbonKit platform, it is vital to guarantee users control over any use or disclosure of their personal information. The CarbonKit should store only service connection information and the user\rq{}s carbon balance, preferences, progress, and goals. Data collected by participating service providers about any one individual must stay within the service provider\rq{}s system and not be stored on the CarbonKit platform. Additionally, personal information should be disclosed only to the specific government agency or private service provider (e.g., Fitbit), and accessed on a need-to-know basis -- records should not be shared across agencies or individual apps.

To further guard privacy, an individual\rq{}s private data, such as credit card numbers and other forms of identification, should be stored on the individual\rq{}s mobile device, and used only to authenticate the individual using the system. Once authenticated, individuals are tracked and data stored based on a unique profile ID. Mobile device such as smartphones with near-field communication apps are a more secure way to authenticate and store transactions within the CarbonKit platform~\cite{mehen2011feedback}.

\section{CarbonKit Incentives}

The CarbonKit would operate on a voluntary basis for personal carbon tracking, offering information for personal use. In subsequent stages, however, the CarbonKit should be able to support more ambitious designs for personal carbon allowances and trading, where individuals who reduced their emissions below allocated levels would sell carbon allowances to those whose emissions exceeded their allocation. People might elect to participate in a voluntary program for many reasons, including:
\begin{itemize}
\item to obtain exemptions, eco-currency or tax rebates;
\item to obtain economic benefits from carbon footprint reductions;
\item interest in the technological features of the system;
\item interest in pursuing goals parallel to carbon footprint reduction, such as health or fitness improvement; and
\item influence by other individuals.
\end{itemize}

Some of these reasons depend on financial support from government or private sector, but the last three reasons need no money as motivation. We envision three types of incentives for the CarbonKit:

\textbf{(1) Opting in to voluntary carbon tracking:} A direct tax rebate combined with in-kind incentives and discounts could improve public adoption and support. Examples of in-kind incentives for carbon tracking include: smartphones to opt into the program and a paid limited monthly usage in exchange for carbon tracking; wireless physical activity trackers; energy efficiency kits; intelligent home monitors; vouchers for home retrofitting; and passes for health and fitness programs.

\textbf{(2) Achieving milestones} Milestones include taking first steps and competing within a circle of influence. Such incentives should be linked with personal goals, and delivered as a reward exchangeable for carbon currency or green services. Governments could finance this type of incentive if they reduce operational costs in areas like health, energy conservation or transportation. Private sponsorship can be an option to fund special campaign incentives.

\textbf{(3) In exchange for actions that benefit others:} This category allows individuals who can neither reduce their carbon consumption nor buy extra allowances to contribute in other ways to reducing emissions. Examples include carpooling programs, hours deposited in a green time bank, and volunteer actions promoted by local governments. This solution can be compared to the concept of offsets in a cap-and trade scheme, since it offers an option to help reduce the emissions of other individuals. However, it would have significant differences -- it promotes the sense of community, and allowances could be obtained only through personal action, not through payment to third persons.

\section{Design Case: BC, Canada}

The CarbonKit as a conceptual model can be applied in many jurisdictions worldwide. However, in order to explain more specific requirements such as the regulatory environment and the applications for each module of the CarbonKit, this section identifies the stakeholders and institutions required to implement a CarbonKit in the Canadian province of British Columbia (BC). A stakeholder network is understood as a group of private and public institutions which collaborate in the design, implementation and operation of a program or project~\cite{affolderbach2012environmental}.
The following discussion provides an overview of those institutions, their competencies, and the unique roles they would play in a CarbonKit. Examples are drawn from BC, but similar stakeholder networks could be identified and constructed for numerous other jurisdictions and countries.
%Table 1 identifies existing applications and services that could be integrated as CarbonKit modules. Many of the apps listed in Table 1 are already in worldwide use (e.g., Facebook) or could be tailored to specific jurisdictions.

\textbf{Provincial/State and Local Governments} always play a leading role in carbon pricing policies. CarbonKit would require the participation of various government institutions working together with the private sector. In BC, some of the core government players would be: the Ministry of Environment as the main authority governing any enabling legislation; the Ministry of Health to collaborate in the implementation of health and fitness improvement goals and to link existing programs and budget with a carbon policy (e.g., My Health, My Community initiative); the Ministry of Technology, Innovation and Citizen Services could collaborate and sponsor the development of the CarbonKit platform and dashboard; the Ministry of Finance would play a leading role in the implementation of a new source of fiscal revenue through the sale of carbon allowances, as well as the correspondent budget to provide incentives. Municipal governments who have policies in place that could be linked to or promoted through CarbonKit (e.g. City of Vancouver Greenest City 2020 Action Plan, City of Surrey Sustainability Charter).

\textbf{Utilities Companies} accumulate knowledge about behavioral change. Energy savings have been required and promoted in BC since the early 1980\rq{}s, long before climate change became a policy concern. BC has set ambitious energy efficiency targets to meet 66\% of new electricity demand through conservation, and to achieve a 20\% reduction in household energy consumption by 2020. Examples of energy companies and programs that could be integrated in the CarbonKit in British Columbia are: LiveSmart BC, BC Hydro\rq{}s Power Smart, and FortisBC\rq{}s PowerSense. These programs provide financial incentives and advice on energy-efficient technologies and practices.

\textbf{Loyalty Management Companies} provide a variety of incentive-based programs. These companies connect databases and transactions with retailers, gasoline stations, banks, air travel, hotels, restaurants, and NGOs, among others. Loyalty management companies have the expertise to design the best type and level of incentives to promote changes in consumer behavior. Companies who could participate in a BC system include: Loyalty One, which operates Air Miles, and AIMIA which operates Aeroplan. Outside BC, the UK-based Social Change Rewards offers points-based incentive programs designed for public sector agencies to reward citizens for making healthier or environmentally beneficial choices.

\textbf{Payment Processors} are a single point of contact that handle transactions from various channels such as credit cards, debit cards, and loyalty. In an operation that usually takes a few seconds, the payment processor will both check the details received by forwarding them to the respective card\rq{}s issuer for verification, and also test the transaction against a series of anti-fraud screens. If verification is denied by the issuer, the payment processor relays the information to the merchant, who then declines the transaction.

\textbf{Social Media} facilitate both access to information and comparison with others, which are important drivers to achieve behavioral change. Companies such as Facebook, Twitter or Klout have revolutionized the world of social interaction, communication and marketing. In many cases, social media can be more effective tools for real-time, multimodal communication that is an effective instrument to promote values like sustainability and social responsibility. Applications to reduce personal carbon footprints have already been developed and delivered through Facebook (e.g., My Sollars offers web/mobile gamified apps for individuals to reduce their carbon footprint and get rewarded for it).

\textbf{Health and Fitness Applications Providers} offer weight management programs often suggest that participants keep a food diary and/or an activity log. Smartphones smart-wareables (e.g., Apple Watch) can provide a vast amount of information to facilitate such a task, from precise calorie calculations to GPS services that calculate the distance covered on a long run. Apps that tracks food intake or total exercise can serve as a reminder to stay the course. What makes mobile apps successful in promoting fitness and health is that people typically have their smartphones them with them at all times. Widely available fitness apps include Fitbit, Misfit, and Moov. These applications can count calories, record exercise activities, and monitoring weight (and other metrics). Many of these applications also use gamification to motivate people to reach a desired goal for exercising.

\textbf{App Developers} can provide innovative apps for a particular cause. In 2010 the BC Government organized an Apps for Climate Action Contest challenging Canadian software developers to raise awareness of climate change and inspire action to reduce carbon pollution by using data in new applications for the web and mobile devices. Winning apps included: Green Money: a personal offset calculator for the money and time people invest in environmental savings; VELO which uses gamification to enable organizations and individuals to monitor and compare their GHG emissions continually rather than annually; and MathTappers: Carbon Choices, an app designed to help students examine the effects of their personal choices on climate change. CarbonKit could offer such a contest which could result in having the best apps integrated as components of the dashboard.

\textbf{Taxes and Accounting Firms} such as Vancouver-based EcoTaxFile, a firm focused on environmental sustainability, provides accountants with the tools to advise their clients on how carbon footprint reduction can also save money. With the carbon calculator on the EcoTaxFile website, people need only the information already required to file taxes. EcoTaxFile exemplifies how accounting services providers could help engage individuals in reducing GHG emissions while saving money.

\textbf{Cleantech Developers} include BC-based energy intelligence companies such as Rainforest Automation who help make homes more efficient. Using a WiFi power sensor and a cloud service with some smart pattern detection algorithms, this company offers a device and software that monitor home\rq{}s electricity and report useful data for saving money on electrical bills. Ecoisme is a similar solution that provides a friendly dashboard for energy usage. Their technology, combining nonintrusive load monitoring and spectrum analysis, can identify home appliances, check their energy efficiency and suggest the best ways to save energy.

\section{Conclusions}

This paper explores how technology, markets and incentives might be combined in ways that empower and reward individuals\rq{} efforts to reduce GHG emissions. It proposes a CarbonKit -- a combination of information sources, computing applications and incentives that would help individuals track and reduce their personal carbon emissions. The CarbonKit is a ubiquitous technology platform that can be implemented with currently available technology to enable individual carbon budgeting and accountability. It combines a goal-and-reward behavioral feedback loop with the Big Data available from tracked activities, apps and social media. It leverages widespread popular experience with smartphone apps and dashboards to track and share their experience within their chosen social circles, and links the long-term and generational benefits of climate change mitigation with near-term personal benefits including health, fitness, economic rewards and social recognition.

Ultimately, personal carbon tracking can promote climate change mitigation to the degree that it promotes individual awareness and results in imagination, engagement and behavioral modification. The CarbonKit platform and dashboard proposed in this paper uses existing technology and platforms, identifies recognizable stakeholder institutions, and entails modest, jurisdiction-specific regulations and safeguards for the protection of privacy and confidentiality.

% The next two lines define the bibliography style to be used, and the bibliography file.
\bibliographystyle{ACM-Reference-Format}
\bibliography{refs}

\end{document}